\documentstyle[12pt,epsfig]{article}

\newcommand{\beq}{\begin{equation}}
\newcommand{\eeq}{\end{equation}}
\newcommand{\bea}{\begin{eqnarray}}
\newcommand{\eea}{\end{eqnarray}}

\def\GeV{{\rm GeV}}
\def\lapprox{\lower .7ex\hbox{$\;\stackrel{\textstyle <}{\sim}\;$}}
\def\gapprox{\lower .7ex\hbox{$\;\stackrel{\textstyle >}{\sim}\;$}}

\begin{document}
\titlepage
\vspace*{0.5cm}
\begin{flushright}
{UR-1623}\\
{December 2000}\\
\end{flushright}
\begin{center}
\vspace*{2cm}
{\Large {\bf Gluon Radiation in Top Quark Production and Decay at 
$e^+e^-$ Colliders}} \\

\vspace*{1.5cm}
 Cosmin Macesanu\footnote{Email address:  mcos@pas.rochester.edu}
and 
Lynne H.~Orr\footnote{Email address:  orr@pas.rochester.edu} \\
\vspace*{0.5cm}
{\it Department of Physics and Astronomy, University of Rochester\\
Rochester, NY~14627-0171, USA\\}

\end{center}

\vspace*{2.cm}
\begin{abstract}
We study  gluon radiation in top quark production above 
threshold at high energy $e^+e^-$ colliders.  We allow for the
top quarks to be off-shell, considering radiation in both the 
top production and decay processes simultaneously.  Our
calculation includes
all top width effects, spin correlations, and $b$ quark mass effects.
We study the effects of gluon radiation on top mass reconstuction and 
examine the interference between production- and decay-stage
radiation, which can be sensitive to the value of the top quark
decay width.

\end{abstract}

\newpage

\section{Introduction}

By virtue of its large mass, the top quark is unique.  
Because it is so heavy, the top quark decays before it can form 
hadrons  \cite{lifetime} and its spin information is passed
along to its decay products.  Top's mass is comparable to the electroweak
symmetry breaking scale, and its Yukawa coupling is suspiciously close 
to unity.  All of this means that top physics will be an important and 
interesting subject in the near future, and the top quark may lead
us to physics beyond the Standard Model.

Future high energy lepton colliders --- $e^+e^-$ and $\mu^+\mu^-$ --- can
provide relatively clean 
environments in which to study top quark physics.  Although top production
cross sections are likely to be lower at these machines than at hadron
colliders, the color-singlet initial states  give lepton machines some
advantages.  Furthermore, the fact that the 
laboratory and hard process center-of-mass frames coincide greatly
simplifies the reconstruction of final states.  In addition, many of 
the top quark's  couplings, 
especially those to the photon and $Z^0$ boson, can be easily studied there.

The potential for precision studies of top  physics at such colliders 
requires precision predictions from the theory, beyond leading order
in perturbation theory.  
In particular, QCD corrections must be taken into account.  
Jets from radiated gluons can be indistinguishable from quark jets, 
complicating identification of top quark events from reconstruction of
top's  decay products.  To make matters worse, 
emission may occur in either the top production or decay
processes, so that radiated gluons may or may not themselves be 
products of the decay.  Subsequent  mass measurements can 
be degraded, not only from misidentification of jets but also
from subtle effects such as jet broadening when gluons are 
emitted near other partons.

In this paper we study the effects of real gluons radiated in
top quark production and decay at $e^+e^-$ colliders.  We consider 
collision energies well above the top pair 
production threshold, so although for definiteness we will refer
to electrons in the initial state, our parton-level results
apply equally well to $\mu^+\mu^-$ collisions at the same energy.
We allow for the top quarks to be off-shell, 
keeping the full width-dependent top propagator and retaining all
spin correlations.
Past treatments of QCD corrections have treated the top 
production \cite{prod} or decay \cite{decay} processes separately, or in one
case \cite{schmidt} combined the two processes, but kept the on-shell 
(narrow-width)
approximation.   Gluon radiation for off-shell top has been
treated previously in the soft gluon approximation  \cite{jikia,kos}.
Here we give an exact treatment for arbitrary gluon energies.
We study properties of the emitted gluons,  top
mass reconstruction, and effects of interference between 
production- and decay-stage gluons that can be sensitive to
the top quark width.
Computation of the virtual QCD corrections for off-shell top 
is currently in progress \cite{mcos} and will be combined with
the present results for a full QCD NLO treatment of top quark 
production and decay.

This paper is organized as follows.  In Section 2 we discuss the 
matrix element calculation and Monte Carlo integration.  We 
present numerical results in Section 3, focusing on gluon 
distributions, mass reconstruction, and interference effects.
We present our conclusions in Section 4. An Appendix includes some comments about 
gauge invariance.

\begin{figure}[t]
\begin{center}
\hspace*{-3cm}\mbox{\epsfig{figure=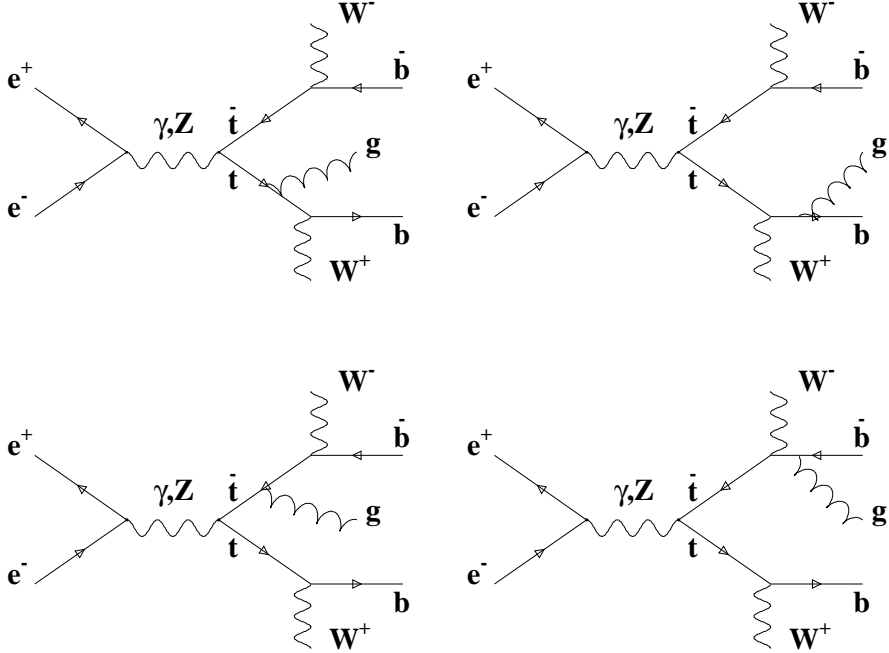,width=11.0cm}}
\caption[]{Feynman diagrams for gluon emission in top production and decay 
at lepton colliders.}
\label{diagrams}
\end{center}
\end{figure}

\section{Cross Section Calculation}

\subsection{Computation of the matrix element}

We compute the cross section for  real
gluon emission in top quark production and decay: 
\begin{equation}
e^+e^- \rightarrow \gamma^*, Z^* \rightarrow t\bar{t} (g)
\rightarrow bW^+ \bar{b}W^-g\; .
\end{equation}
At $e^+e^-$ colliders,  no gluons are radiated from the 
color-singlet initial state.  Final-state gluon emission can occur in 
both the production and decay processes, with gluons emitted from the
top or bottom quarks (or antiquarks), as shown in Figure~\ref{diagrams}.
Emission from the top quark contributes to both production- and decay-stage
radiation, depending on when the top quark goes on shell.  Emission 
from the $b$ quarks contributes to decay-stage radiation only.  The
separation of these contributions will be discussed below.

We compute the exact matrix elements for the diagrams shown in 
Fig.~\ref{diagrams} with all spin correlations and the bottom mass
included,  using the helicity methods
of Kleiss and Stirling     \cite{ks}.  
Working at the matrix element, rather than
 the matrix element squared, level has the usual advantages of 
numerical efficiency, and in our case has the additional 
advantage that we can identify individual contributions and 
interferences between them.  The explicit 
expressions for the matrix elements are complicated and not particularly 
illuminating, so we do not reproduce them here\footnote{A FORTRAN 
program containing the matrix elements  
can be obtained from the authors.}.

We do not assume the top quark to be 
on shell; therefore we keep the 
finite top width $\Gamma_t$ in the top quark propagator and include
all interferences between diagrams.  We use exact kinematics in
all parts of the calculation.
We do not include radiation from hadronic $W$ decays; the 
$W$ bosons are assumed to decay leptonically and we integrate over 
their decay products in the results presented here. 
In practice radiative hadronic $W$ decays could be largely
eliminated through invariant mass cuts, but a detailed study is
required to determine exactly how effective this would be.

Some comments about gauge invariance are in order here.  In 
the on-shell, or narrow-width, approximation, the top quarks are
always on mass shell and the width is identically zero.  In that case
gauge invariance requires only diagrams with two intermediate
top quarks.  Once the top quarks are allowed to have a finite width and
be off-shell, however, nonresonant diagrams --- those with the same 
final state but only one or no intermediate top quarks ---  
can also contribute.  In the region of interest for top physics,
{\it viz.,} the region where the top quarks are nearly on shell,
the poles in the doubly-resonant diagrams (those with two tops)
cause their contributions to dominate.  This is the motivation
for the ``double-pole approximation'' in electroweak radiative corrections
to $W$ pair production     \cite{WW}.  In the present case we use 
invariant mass cuts (see below) to guarantee that
 we are in a region of the phase space where the doubly
resonant diagrams  dominate the cross section. 
We can restore gauge invariance using the approximation described in the 
appendix.  This gauge invariant result differs
from the exact one by non-doubly-resonant
terms, and this difference is very small numerically in the
regions of interest.

\subsection{Production--decay decomposition}

As mentioned above, calculating at the amplitude level allows us to 
identify contributions from individual processes and 
their interferences.  We are particularly interested in 
distinguishing between contributions from gluons
radiated in the top quark production and decay stages.  
This is directly related to reconstruction of the top quark
momentum  from its decay products,
which in an experiment 
allows us both to identify top events and to measure $m_t$. 
 The presence of gluon radiation complicates 
the reconstruction because 
the emitted gluon may or may not be part of the top decay.
If the gluon is not part of the decay, then it is represents
a correction to top production and  should not be included in
the top momentum reconstruction:
\beq
m_t^2\approx p_t^2=(p_b+p_W)^2\equiv p_{bW}^2 \; .
\label{mtprod}
\eeq
If on the other hand the gluon is part of the decay, then it 
{\it should} be  included in top reconstruction:
\beq
m_t^2\approx p_t^2=(p_b+p_W+p_g)^2\equiv p_{bWg}^2 \; . 
\label{mtdecay}
\eeq
Being able to make this distinction turns out to be useful for purposes
of efficient phase-space integration as well.

Although this production--decay distinction cannot be made absolutely 
in an experiment\footnote{If the interference between processes is 
large this distinction is not even meaningful.}, 
the various contributions can be separated in the calculation.  
For radiation from the $b$ and $\bar{b}$ quarks, the assignment is 
easy:  these contributions, corresponding to the two right-hand 
diagrams in Fig.~\ref{diagrams}, are clearly part of the top
quark decay.  However, as noted
above,  gluon emission from the top quark (or antiquark) contributes
to both the production and decay stages; which is which depends on 
whether the  top was closer to its mass shell  before or after  
emitting the gluon.  This condition corresponds to which of the two 
propagators from the top that emitted the gluon is numerically larger.

We can make the separation in our calculation
as follows \cite{kos}.  For definiteness, we consider gluon emission from
the top quark, shown in the upper left diagram in Fig.~\ref{diagrams}.
The matrix element for this diagram contains propagators for the 
top quark both before 
and after it radiates the gluon.  The matrix element therefore 
contains the factors 
\begin{equation}
{\cal{M}} \, \propto \,\left( {{1}\over{p_{Wbg}^2-m_t^2+im_t\Gamma_t}}\right)
\left( {{1}\over{p_{Wb}^2-m_t^2+im_t\Gamma_t}}\right)\; .
\label{twoprop}
\end{equation}
The right-hand side can be rearranged to give 
\begin{equation}
{\cal{M}}\, \propto \,{{1}\over{2p_{Wb}*p_{Wbg}}}
\left( {{1}\over{p_{Wb}^2-m_t^2+im_t\Gamma_t}} -
{{1}\over{p_{Wbg}^2-m_t^2+im_t\Gamma_t}}\right)\; .
\label{decomp}
\end{equation}
This separates the production and decay contributions to the matrix element.
The first term in parentheses contains a propagator that 
peaks when  $p_{Wb}^2=m_t^2$, which corresponds to the condition for
production stage, as in Eq.~\ref{mtprod}.  Similarly, the 
second term peaks for $p_{Wbg}^2=m_t^2$, which 
corresponds to decay emission as in Eq.~\ref{mtdecay}.  

The complete amplitude can now be written schematically as 
\beq
{\cal{M}}_{tot} = {\cal{M}}_{prod} + {\cal{M}}_{tdecay}+ {\cal{M}}_{\bar{t}decay}\; .
\label{mtot}
\eeq
The cross section, obtained from taking the absolute square of 
${\cal{M}}_{tot}$,
then contains separate production and decay contributions, from 
$|{{\cal{M}}}_{prod}|^2$ and  $|{\cal{M}}_{tdecay}|^2, |{\cal{M}}_{\bar{t}decay}|^2$,
respectively.  It also contains cross terms representing the interferences,
which in principle confound the separation but in 
practice are quite small.

The interference terms {\it are}  interesting in their own right, although
not for top reconstruction.  In particular, the interference between 
production- and decay-stage radiation can be sensitive to the top
quark width $\Gamma_t$  \cite{jikia,kos}, which is  1.42 GeV in the 
Standard Model at ${\cal{O}}(\alpha_s)$  \cite{decay}.
The interference between for example the two propagators shown 
in Eq.~\ref{decomp}  can be 
thought of as giving rise to two overlapping Breit-Wigner resonances.  The
peaks are separated roughly by the gluon energy, and each curve
has width $\Gamma_t$.  Therefore when the gluon energy becomes comparable 
to the top
width, the two Breit-Wigners overlap and there can be substantial
interference.
In constrast, if the gluon energy is much larger than $\Gamma_t$, the 
overlap --- and hence the interference ---  is negligible. 
Therefore  the amount of interference 
serves as a measure of the top width.  We will explore this more below.

The integration over the final state phase space to obtain the cross
section involves an integrand that contains multiple Breit-Wigner
peaks from the top quark propagators as well as infrared singularities
when the gluon energy becomes small.  Even with 
cuts on $E_g$, the rapid variation of the integrand can spoil the integration
procedure. To eliminate this problem, we tailor the momentum generator
to the production of a gluon in association with two massive particles
($\gamma^*, Z^* \rightarrow t \bar{t} g$ or $t  \rightarrow b W g $).
The multiple Breit-Wigner peaks are taken into account by using a 
multi-channel approach that integrates separately over the 
individual production and decay contributions; the Breit-Wigner
behavior is smoothed out in the phase space generation.  The 
interference terms, which have products of Breit-Wigners that
peak in different places, much like in Eq.~\ref{twoprop}, are
 are integrated using
a combination of the three main channels.

\section{Numerical Results}

In this section we show results of the numerical calculation described
above.  We present the cross section for 
$b\bar{b}W^+W^-g$ production in
$e^+e^-$ collisions at a 500 GeV center-of-mass 
energy, with a few exceptions which are clearly identified.  
The calculation is entirely at the parton level, and we do not 
include initial state radiation, beam energy spread, or beamstrahlung.
We use the 
following numerical values of parameters:  $m_t=175\ \GeV$,
$m_b=5\ \GeV$, $M_W=80\ \GeV$, $\Gamma_t=1.42\ \GeV$, and $\alpha_s=0.1$.
Note that $\alpha_s$ appears simply as an overall factor, because
all of our events contain a gluon.  

Unless otherwise indicated, we use the following cuts.  We require
$E_g>5\ \GeV$ to eliminate the infrared singularity and because
we intend for the gluon to be detectable.  In addition we wish
the gluon to be separable from the $b$ and $\bar{b}$ quark;
this is implemented via the requirement $m_{bg}, m_{\bar{b}g}> 10\ \GeV$,
which we shall identify below as ``$m_{bg}$ cuts.''
(Separation could also be achieved with a cut on the gluon's $E_T$
with respect to the $b$ or $\bar{b}$;  the choice
makes little difference in the resulting distributions.)
In order to
 make sure that we do not get contributions to our results
from regions of the phase space where non-doubly-resonant diagrams 
might be important, we require
\beq
160\ \GeV \,\leq \,m_{bW}\, \leq \,190\ \GeV \;\;\;\;\; {\rm or} \;\;\;\;\;
160\ \GeV \,\leq \,m_{bWg}\, \leq \, 190\ \GeV 
\eeq
and the same thing for the $\bar{b}$.  These conditions will be identified
as ``$m_t$ cuts'' below.

\begin{figure}[ht]	
\vskip -.25 cm
\hspace*{2cm}
\mbox{\epsfig{figure=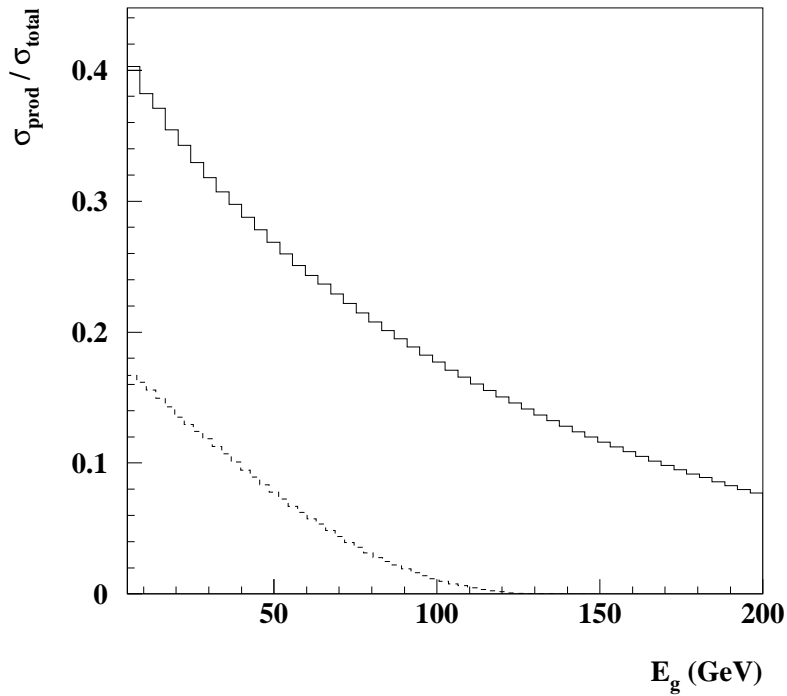,width=13.0cm}}
\caption[]{
\label{prodfrac}
\small The fraction of gluon emissions radiated in the production stage, as
a function of minimum gluon energy, for center-of-mass energy 1 TeV (solid
line) and 500 GeV (dashed line), with no cuts.}
\end{figure}
 
\subsection{Characteristics of the gluon radiation}

We begin  with the  relative contributions 
of production-
and decay-stage radiation to the total cross section.  Figure \ref{prodfrac}
shows the fraction of the total cross section due to production stage 
emission in events with an extra gluon, as a function of the 
minimum energy of the gluon.   This figure contains no cuts 
besides that for gluon energy and is simply meant to
illustrate how radiation is apportioned in top
production and decay for different center-of-mass energies; 
the solid line corresponds to c.m.\ 
energy 1 TeV, and the dashed line is for 500 GeV.  
Both curves fall off 
as the minimum gluon energy increases; this reflects the decrease in 
phase space for gluons radiated in the production stage.  We see that 
 the production fraction is always higher at 
1 TeV collision energy than at 500 GeV.  This too reflects phase
space --- for a given gluon energy there is more phase space
available to produce gluons in association with top
pairs  at the higher c.m.\ energy.
However both fractions remain below 0.5; decay-stage radiation always 
dominates at these energies.

\begin{figure}[ht]	
\mbox{\epsfig{figure=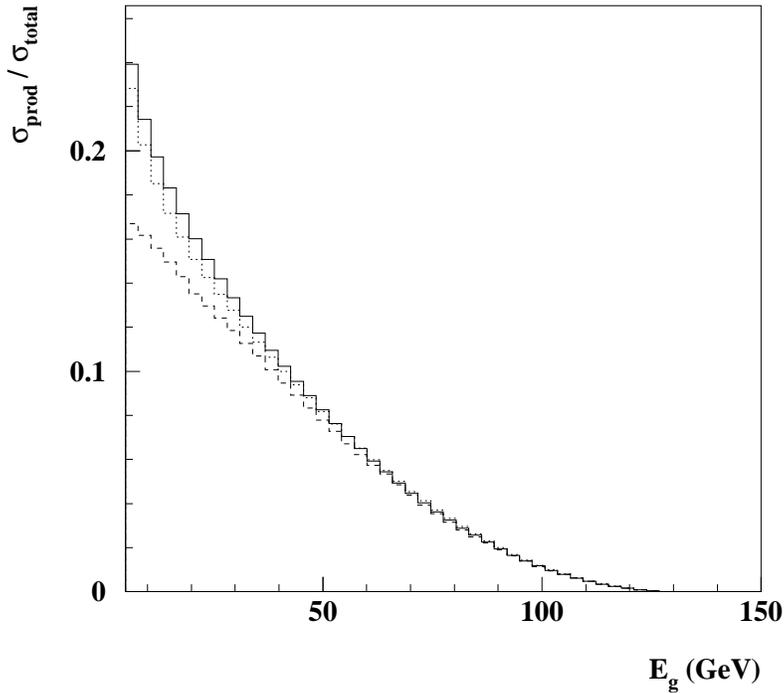,width=13.0cm}}
\vskip -.5 cm
\caption[]{
\small The fraction of gluon emissions radiated in the production stage, as
a function of minimum gluon energy, for center-of-mass energy  500 GeV,
with no cuts (dashed line), $E_T(g,b)>3\ \GeV$ (dotted line),
and $m_{bg}>10\ \GeV$ (solid line).}
\label{prodfraccuts}\end{figure}

Figure \ref{prodfraccuts} shows for a 500 GeV center-of-mass
energy the effect on the production fraction
of separation cuts between the gluon and $b$ quarks.  The dashed line
shows the fraction with no cuts.  The dotted line corresponds to
requiring that the transverse energy of the gluon with respect to 
the $b$ and $\bar{b}$ --- which we denote $E_T(g,b)$ ---
be greater than 3 GeV.  The solid line corresponds to the 
cut $m_{bg}>10\ \GeV$ where the $b$ can either be a quark or antiquark.
The effect of both of these cuts is to eliminate gluons that are
soft and/or close to one of the bottom quarks; since these
contributions tend to come from decay-stage radiation, their effect
is to increase the fraction of production-stage radiation. If the
$b$ were massless there would be a collinear singularity in the decay
contribution; this does
not happen in our case but the decay distribution still peaks when
the $b$-quark--gluon angle is small.   
The effects of both cuts become smaller
with increasing gluon energy.

\begin{figure}[ht]	
\mbox{\epsfig{figure=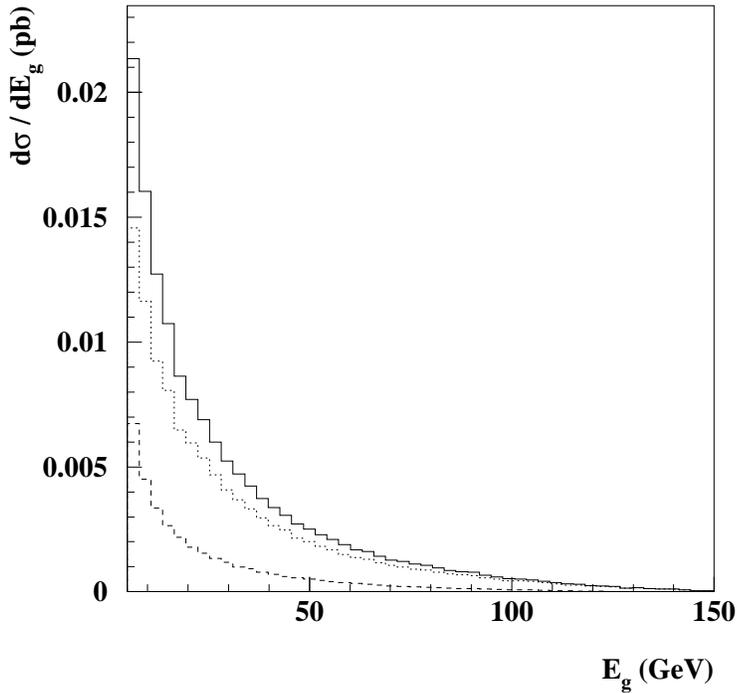,width=13.0cm}}
\vskip -.75 cm
\caption[]{
\label{energy}
\small The  spectrum of radiated gluons as a function of gluon energy in
GeV for center-of-mass energy 500 GeV, with $m_{bg}$ and $m_t$ cuts
(see text).  Dashed histogram:  production-stage radiation.  Dotted histogram: 
decay-stage radiation.  Solid histogram:  total.}
\end{figure}

Figure \ref{energy} shows the total gluon energy spectrum for a
collision energy of 500 GeV along with its decomposition into production
(dashed histogram) and decay (dotted histogram) contributions. 
The interferences between the two are negligible and are not shown; this 
will be true for all subsequent figures until  we consider the 
interference explicitly.  Included in this figure are the $m_t$ and 
$m_{bg}$ cuts discussed above.  As indicated in the previous figures, 
radiation from the top decays dominates.  
Otherwise the spectra are not vastly 
different; both exhibit the rise at low energies due to the 
infrared singularity characteristic of gluon emission, and both fall
off at high energies as phase space runs out.

\begin{figure}[ht]		
\mbox{\epsfig{figure=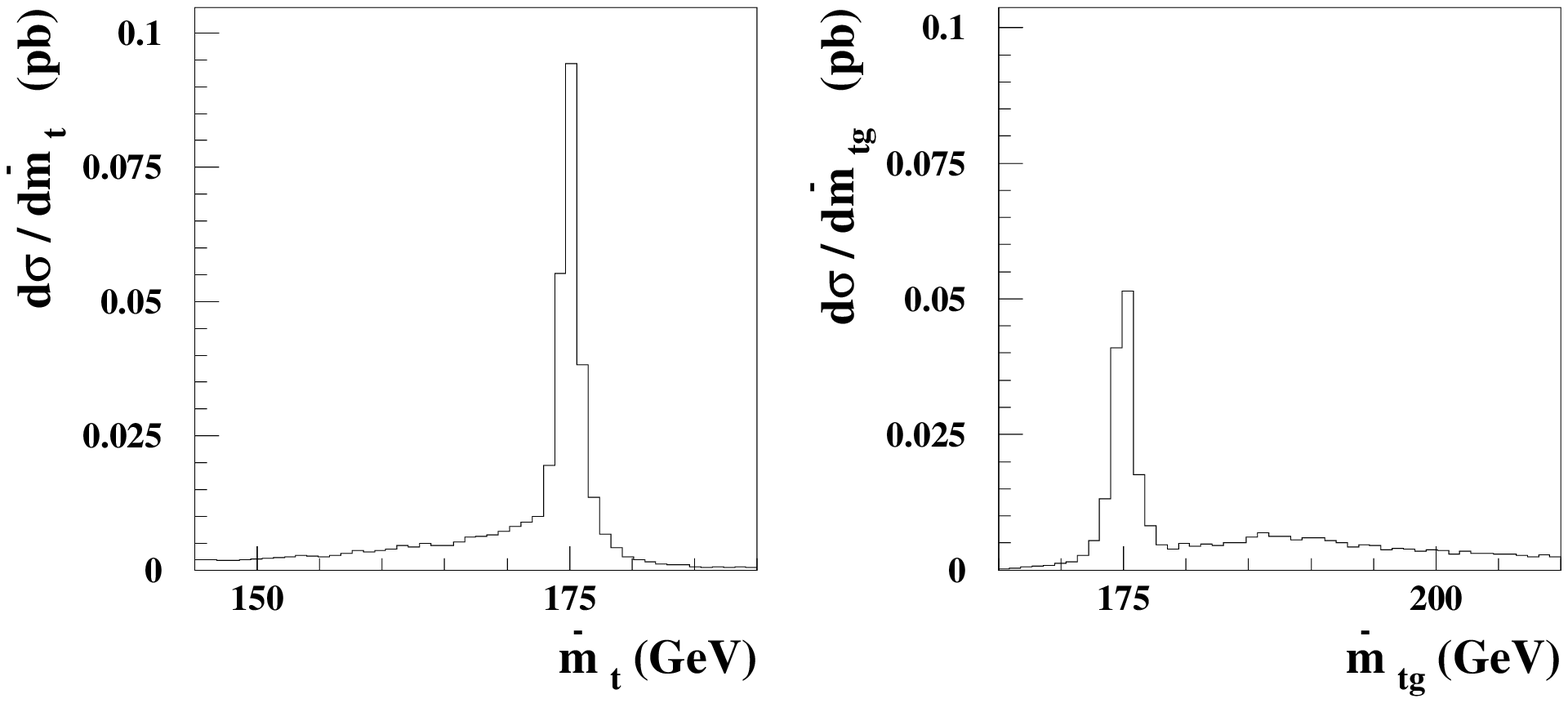,width=13.0cm}}
\vskip -6cm
\caption[]{
\label{masses}
\small The  top invariant mass spectrum without (left) and with (right) 
the gluon momentum included, for center-of-mass energy 500 GeV, with
$m_{bg}$ cuts and $E_g>5 \GeV$.}
\end{figure}

\subsection{Mass Reconstruction}

We now turn to the question of mass reconstruction.  
Figure \ref{masses} shows top invariant mass distributions with and without
the extra gluon included; the first plot shows the distribution in 
$m_{bW}$ and the second shows $m_{bWg}$.  We have imposed $m_{bg}$ cuts
and required $E_g>5\ \GeV$.  
  In both cases there is a clear peak  at the 
correct value of $m_t$.  Note that the peak in the 
first plot contains the production contribution as expected,
but the radiative decay part contributes as well.  
This is because even for decay-stage radiation, only one of 
the produced $t$ quarks decays radiatively; the other still has 
$p_t^2=p_{bW}^2$ and therefore contributes to the $m_{bW}$ peak.
The long tails in the two distributions are from misassignments of
the gluons.  
In the left-hand plot, where the gluon
is not included in the reconstruction, we see a low-side tail due to 
events where the gluon was radiated in the decay but was not included
in the reconstruction.  Similarly, 
in the right-hand plot we see a high-side tail due to events where the 
gluon was radiated in association with production, and was included when it 
should not have been.

\begin{figure}[ht]		
\mbox{\epsfig{figure=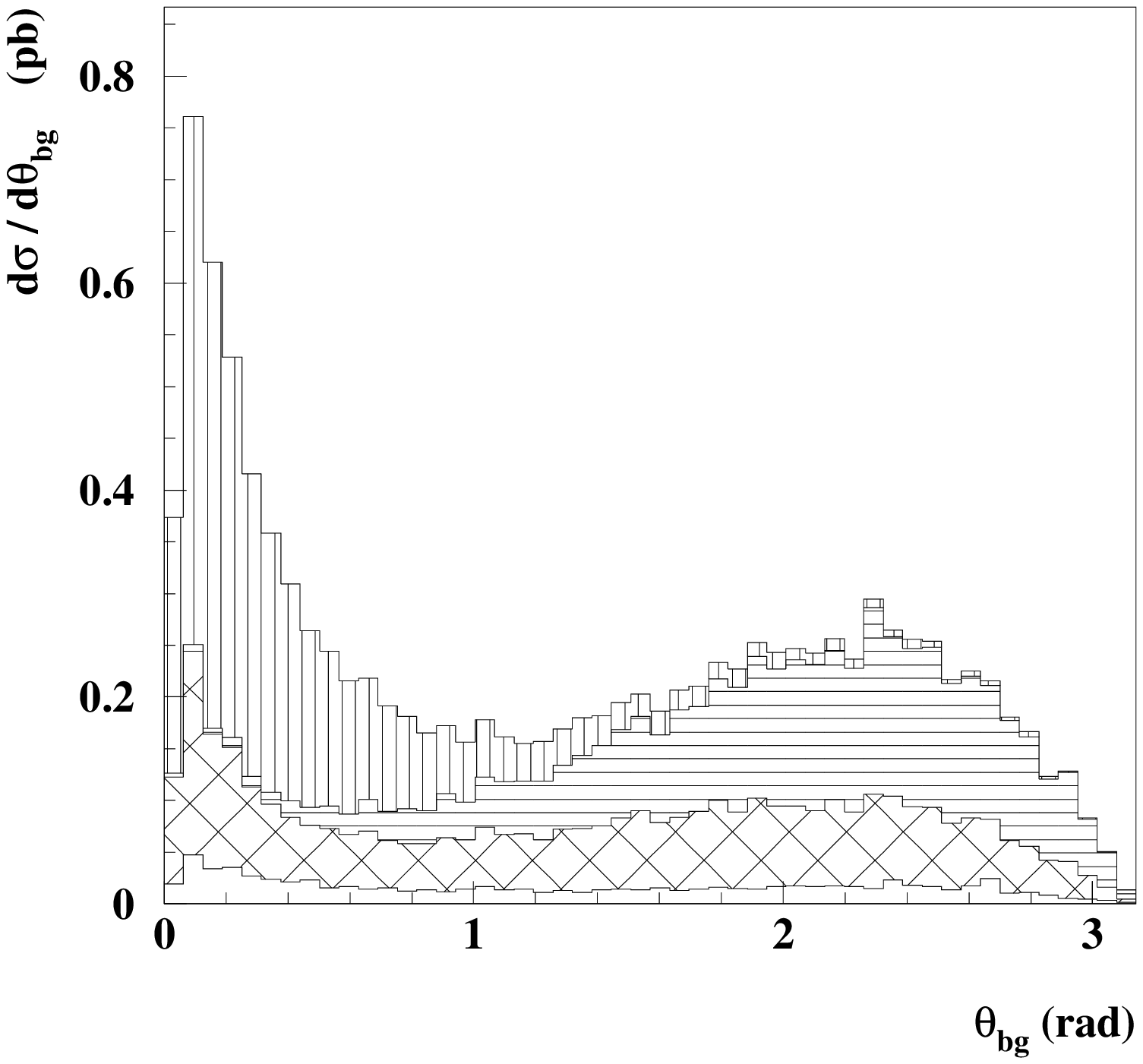,width=13.0cm}}
\vskip -1cm
\caption[]{
\label{bgangle}
\small The  distribution in the angle between the 
 gluon and the $b$ quark for center-of-mass energy 500 GeV, with
$E_g>5\ \GeV$. The various contributions are as described in the text.}
\end{figure}

We wish to define a single distribution for the top mass that 
combines both types of events yet omits wrong combinations as much 
as possible.  One possibility is to cut on the angle between
the gluon and the $b$ quark, whose distribution we show in
Figure \ref{bgangle}.  This is motivated by the fact that 
gluons radiated from the $b$ quarks are always part of the 
decay, and such gluons tend to be emitted close to the 
$b$ quark direction.  
As we have mentioned, the mass
of the $b$ quark prevents a collinear singularity, but the
gluon distribution still peaks close to the $b$, as can be 
seen in the figure.  Because we wish to define cuts that 
give a narrow top invariant mass distribution, the
distribution in $\theta_{bg}$ is decomposed into various
invariant mass regions.  (Here we refer to the $b$ quark only,
and not the $\bar{b}$.)    
 Using the variables $\tilde{m}_t=m_{bW^+}$, 
 $\tilde{m}_{tg}=m_{bW^+g}$, $\tilde{m}_{\bar t}=m_{\bar{b}W^-}$ and
 $\tilde{m}_{{\bar t}g}=m_{\bar{b}W^-g}$  
  we define four types of events: \\
\hspace*{0.5in} type 1 : $172\ \GeV < \tilde{m}_{tg} , \tilde{m}_{\bar{t}} <178\ \GeV$ 
(vertical hatching)\\
\hspace*{0.5in} type 2 : $172\ \GeV < \tilde{m}_{t} , \tilde{m}_{\bar{t}g} <178\ \GeV$ 
(horizontal hatching)\\
\hspace*{0.5in} type 3 : $172\ \GeV < \tilde{m}_{t} , \tilde{m}_{\bar{t}} <178\ \GeV$ 
(cross hatching)\\
    \hspace*{0.5in} type 4 : any other event  
(no hatching)\\
Type 1 events are dominated by contributions from radiative $t$ decays,
and we can see that they do tend towards the $b$ quark
direction.  
Type 2 events (horizontal hatching) are in turn
dominated by radiative $\bar{t}$ decays; gluons in
this case tend to cluster near the $\bar{b}$ direction, and since
the $b$ and $\bar{b}$ tend to appear in opposite hemispheres,
type 2 gluons are mostly found at large angles to the $b$.  
Events of type 3 (cross hatching) are 
mostly production-stage contributions; their distribution is more or less
uniform, independent of the $b$ quark direction.
Finally, events of type 4 (no hatching)
get contributions from both production and decay, with no compelling 
evidence for one over the other.

\begin{figure}[ht]		
\mbox{\epsfig{figure=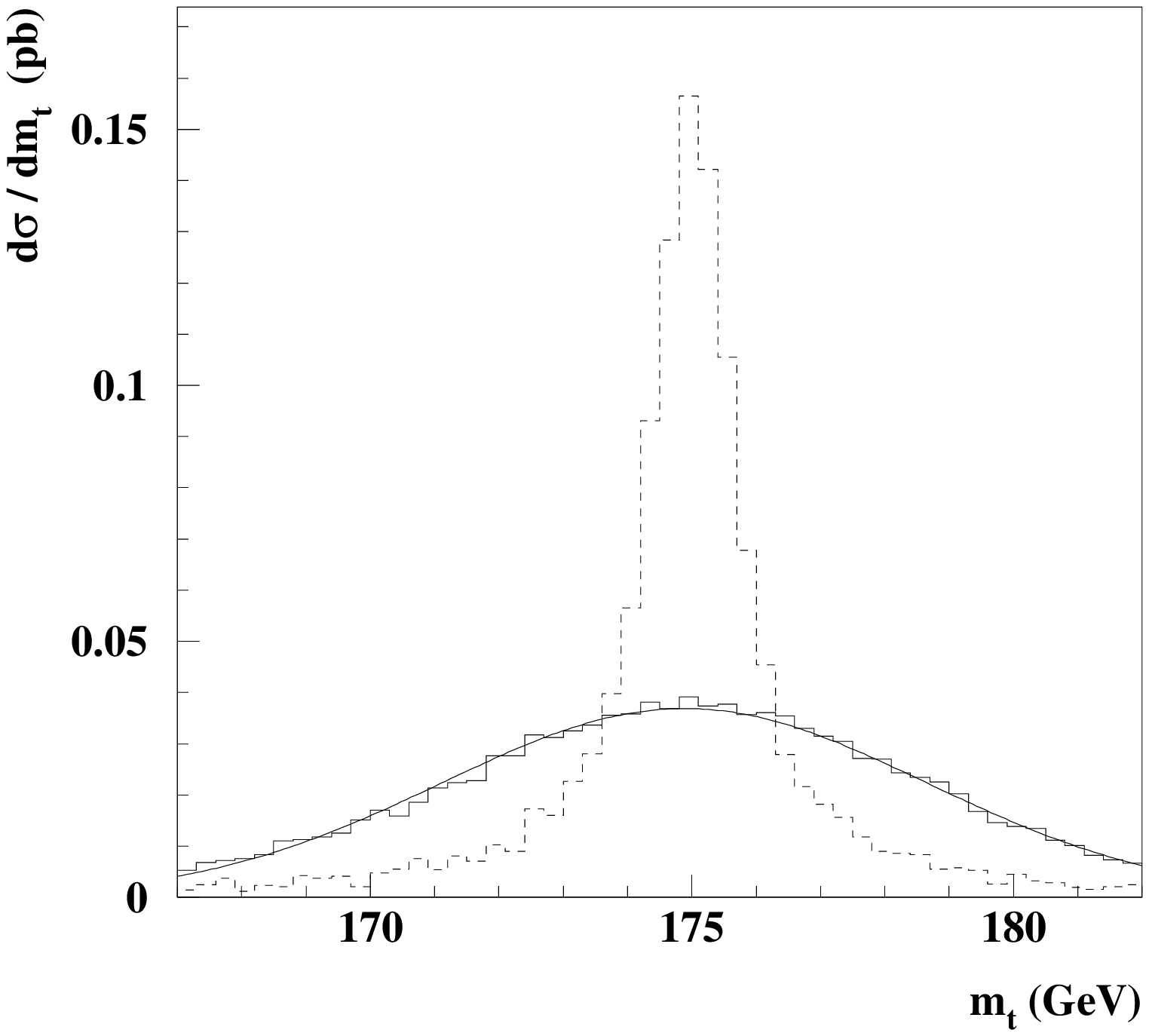,width=13.0cm}}
\vskip -1.25cm
\caption[]{
\label{masssmear}
\small The  top invariant mass spectrum with $b$-gluon angle selection
criteria (dotted histogram), for center-of-mass energy 500 GeV,
minimum gluon energy 5 GeV, and $m_{bg}$ cuts.  
The solid curve and histogram show the effects of energy smearing.}
\end{figure}

Using this figure we can make the following conventions:\\
\hspace*{0.5in} if $\theta _{bg} < 0.7 \ {\rm rad}, $ assign gluon to $t$ decay \\
 \hspace*{0.5in} if $\theta_{\bar{b}g} < 0.7 \ {\rm rad},$ assign gluon to $\bar{t}$
  decay \\
\hspace*{0.5in} if $\theta_{bg}, \theta_{\bar{b}g} >1\ {\rm rad}, $ 
 assign gluon to $t\bar{t}$ production.\\
 With these cuts on the proximity of the gluon to the 
$b$ quark, we construct  the top mass distribution  presented
in the dotted histogram in Figure \ref{masssmear}.

Of course an important reason the cuts are so effective is that we 
work at the 
parton level.  The experimentalists do not have that luxury, and, as one
would expect, hadronization and detector effects are likely to cloud
the picture.  The solid histogram in Fig.~\ref{masssmear} shows the
mass distribution after including energy smearing; the solid curve is a 
Breit-Wigner fit.   
The spread in the measured momenta of the final state particles is 
parametrized by Gaussians with
widths $\sigma=0.4 \sqrt{E}$ for quarks and gluon, and $ 
\sigma=0.15 \sqrt{E}$ for the $W$'s.  We see that the central value does not 
shift, but the distribution becomes significantly wider.  

These results are meant to give an indication of the effects of hard gluon
radiation 
on mass reconstruction and how they might be dealt with.  Other variables
to consider in choosing the cuts are $m_{bg}$,  the transverse energy
of the gluon with respect to the $b$ or $\bar{b}$, or some combination
of energies and angles as defined in the various algorithms used
in jet definitions for $e^+e^-$ colliders.    At tree
level and with partons only, the exact choice is not very important.
We will revisit the question in more detail when we include virtual
corrections in a full NLO calculation.

\begin{figure}[ht]		 
\mbox{\epsfig{figure=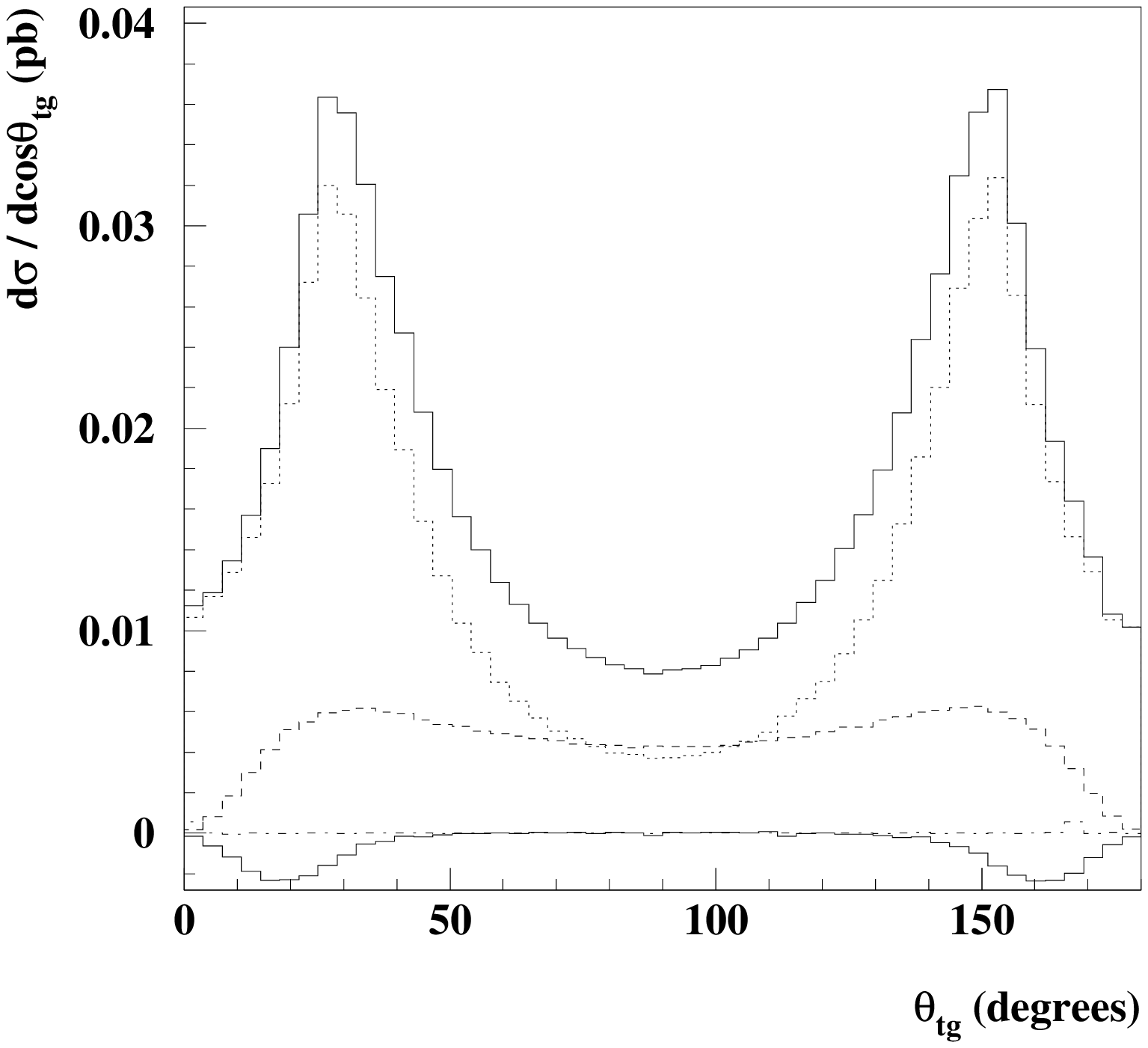,width=13.0cm}}
\vskip -1 cm
\caption[]{
\label{thetagt}
\small The  distribution in angle between the top quark and the gluon for 
gluon energies from 5 to 10 GeV, 
$\cos\theta_{tb},\cos\theta_{\bar{t}\bar{b}}<0.9$, $m_t$ cuts,
 and 750 GeV collision
energy.  The upper solid histogram is the  total and the other histograms
represent the individual contributions:  dotted:  decay; dashed: production;
dot-dashed: decay-decay interference; solid:  production-decay interference.}
\end{figure}

\subsection{Interference and Sensitivity to $\Gamma_t$}

Finally, we return to the subject of interference.  As mentioned above, the 
interference between the production- and decay-stage radiation can 
be substantial for gluon energies close to the 
total width of the top quark $\Gamma_t$; the interference 
is therefore  sensitive to the value of $\Gamma_t$.  
However, because this interference is in general small, we need to find
regions of phase space where it is enhanced.  This question was considered
in  \cite{kos} in the soft gluon approximation\footnote{See also 
 \cite{jikia,siopsis}}, where it was found that 
the interference was enhanced when there was a large angular separation
between the $t$ quarks and their daughter $b$'s.

Here we examine whether
the result of \cite{kos}, which considered a  fixed final-state
configuration, 
survives the exact calculation and phase space integration.  
Figure \ref{thetagt}
shows that it does.  There we plot the distribution in the angle between 
the 
emitted gluon and the top quark for gluon energies between 5 and 10 
GeV and with $\cos\theta_{tb}<0.9$ and $m_t$ cuts.  
The center-of-mass energy is 750 GeV. This c.m.\ energy is chosen because
for there to be significant interference between production
and decay-stage radiation, both contributions must be sizable.  
At 500 GeV, we see from Figs.~\ref{prodfrac} and \ref{prodfraccuts}
that the production contribution is suppressed compared to that from decay;
as a result, the interference is very small.  Increasing the 
energy increases the production-stage contribution.  We note that 
the distributions at 750 GeV and 1 TeV do not differ substantially.

The  histograms in Fig.~\ref{thetagt} 
show the decomposition into the various contributions. The  production-stage
radiation is shown as a dashed histogram; we see that it reaches its
larges values at relatively small and large angles.  Small angles
correspond to the $t$ direction, and large angles more or less to the 
$\bar{t}$ direction, since for the small gluon energies of interest here, the $t$ 
and $\bar{t}$ are nearly back-to-back.  The dotted histogram represents
the decay-stage contribution; it dominates the cross section and 
 peaks in the same region as the production contribution.  This leads
to substantial production-decay interference, shown as the negative solid
histogram.  This interference is destructive, so that it serves to 
{\it suppress} the total cross section, shown as the positive solid
histogram.  This effect would be enhanced if
we lowered the gluon energies to values closer to $\Gamma_t$, but 
jets from very low energy gluons are not likely to be observable, so we 
cut off the gluon energy at 5 GeV.  Finally, interference
between the emissions in the $t$ and $\bar{t}$ decays are shown
as a dot-dashed histogram, but as there is very little overlap between 
the two phase space regions even with these cuts, this contribution is 
negligible.

\begin{figure}[ht]		 
\mbox{\epsfig{figure=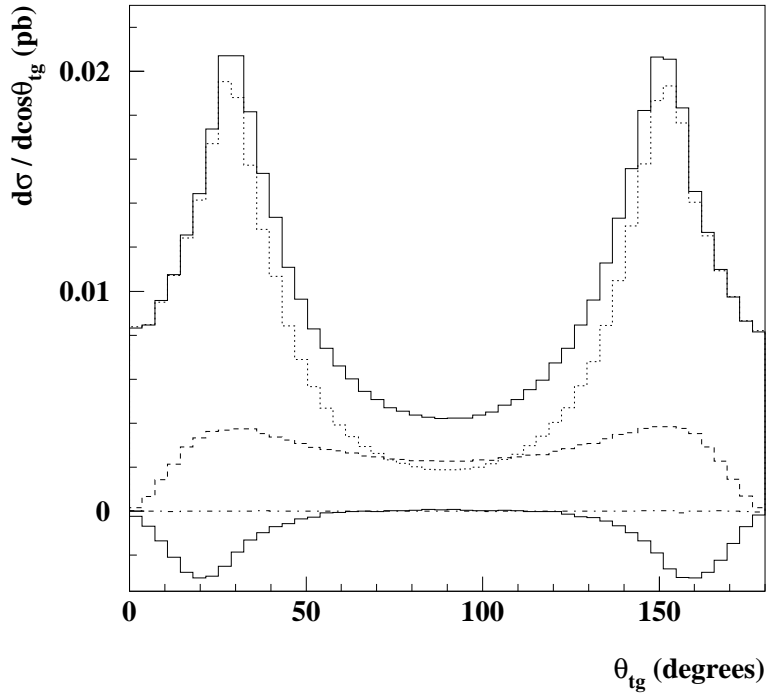,width=13.0cm}}
\vskip -1 cm
\caption[]{
\label{thetagtcuts}
\small As in Fig.~\ref{thetagt}, with the addition of the cuts given
in Eqs.~\ref{interferencecutsa},\ref{interferencecutsb}.}
\end{figure}

\begin{figure}[ht]		
\mbox{\epsfig{figure=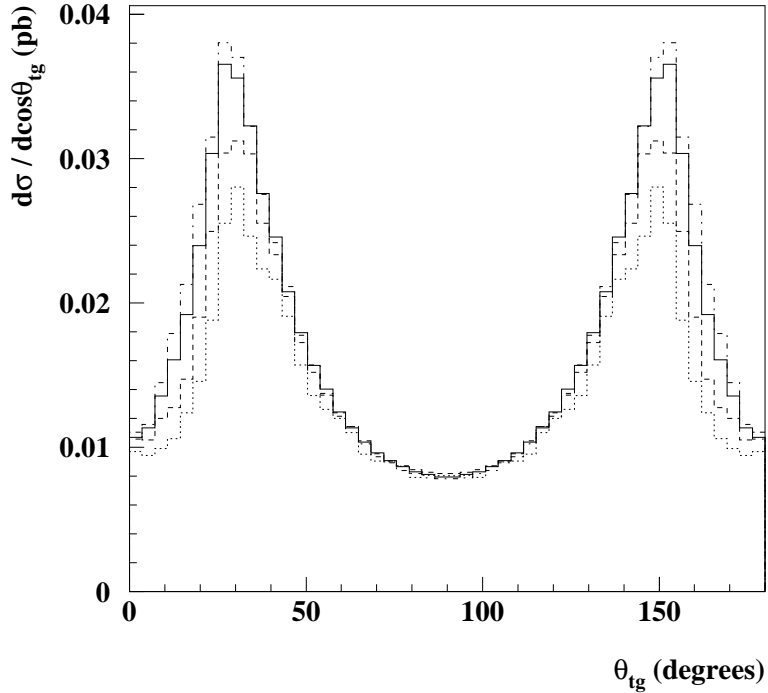,width=13.0cm}}
\vskip -1 cm
\caption[]{
\label{width}
\small The  distribution in angle between the top quark and the gluon for 
gluon energies from 5 to 10 GeV, 
$\cos\theta_{tb},\cos\theta_{\bar{t}\bar{b}} <0.9$, 
and 750 GeV collision
energy.  The histograms correspond to different values of the top 
width $\Gamma_t$:  dot-dashed: 0.1 GeV; solid: 1.42 GeV (SM); 
dashed: 5. GeV;
dotted: 20 GeV.}
\end{figure}

The cuts we have used are fairly generic; we can further enhance
the interference terms with a judicious choice of additional cuts. 
If we 
examine their  behavior in more detail in various
regions of phase space, we find that 
the sign of the interference terms depends on the value of
invariant mass of the top quark.    Since we  integrate
over this  mass, we get cancellations (a similar effect
ensures cancellations of non-factorizable corrections in inclusive 
quantities).

Consider the interference between radiation in the production stage and 
the top decay
stage. The product of the two Breit-Wigner peaks is proportional to the  factor
\beq 
f_{t} = (p_{Wb}^2 - m_t^2)*(p_{Wbg}^2 - m_t^2) + m_t^2 \Gamma_t^2 
\eeq
This factor will multiply a quantity which, upon integration over angles,
is negative. Therefore, for invariant mass values such that $f_{t}$ is 
positive, the interference terms are negative, while 
for negative $f_{t}$,  the interference terms are positive.
We can impose cuts that take advantage of this:  if we require
the invariant masses to satisfy
\beq
f_{t} > 0\ , \ \mbox{if}\ \theta_{bg} <  \theta_{\bar{b}g} ; 
\label{interferencecutsa}
\eeq
\beq
 f_{\bar{t}} > 0\ , \ \mbox{if}\ \theta_{bg} >  \theta_{\bar{b}g}\; , 
\label{interferencecutsb}
\eeq
we obtain the distribution shown in Figure \ref{thetagtcuts}.  The 
interference effects are enhanced, though at the cost of a substantial decrease 
in cross section.

Because the production-decay interference is destructive, increasing
the top width would further suppress the total distribution.  The height
of the peaks, then, is sensitive to the value of $\Gamma_t$.  
This is illustrated in Figure \ref{width}, which
shows the cross section (without the cuts of 
Eqs.~\ref{interferencecutsa},\ref{interferencecutsb}) 
for different values of the top 
width\footnote{The histograms here are scaled so that they all would have
the same normalization
in the absence of interference effects.  Without this rescaling, changing
the width while keeping the $tbW$ coupling fixed changes the total
cross section, which behaves like $1/\Gamma_t^2$ for small $\Gamma_t$.}, 
ranging from 0.1 GeV to 20 GeV.  The SM case ($\Gamma_t=1.42\ \GeV$) is
shown as a  solid line.  It is interesting to note that 
in the context of perturbative gluon radiation, the SM top width
is actually a small quantity.   There are several other points to note.
In principle, this sensitivity to $\Gamma_t$ gives us
a method to measure the top quark's total  width,
independent of decay mode, above the top production threshold.
Although in practice statistics would surely limit the 
possible precision of such a measurement, the total top width is not
so easy to measure directly by any means.  Furthermore, the effects 
illustrated here arise from simple quantum-mechanical interference,
and finding experimental evidence for interference between the 
radiation at the various stages is an interesting goal by itself.

\section{Conclusion}

In summary, we have presented results from an exact
parton-level calculation of hard gluon radiation in off-shell
top production and decay above threshold
at lepton colliders, with the $b$ quark mass and finite top width, 
as well
as all spin correlations and interferences included.  
We have decomposed the cross section into the separate contributions
from emissions associated with top production, $t$ and $\bar{t}$ decay,
and their interferences.  
We have indicated
some of the issues associated with this gluon radiation in top mass
reconstruction and top width sensitivity in the gluon distribution.  
A detailed treatment of many of these issues, in particular with regard to 
mass reconstruction, will be presented in forthcoming work which
combines virtual and real QCD corrections to this process into
a complete NLO computation of top production and decay.

\vspace{1.0cm}
\noindent {\large \bf Acknowledgements} \\

\noindent 
We thank C.R.~Schmidt and W.J.~Stirling for helpful correspondence and 
discussions.  
This work was supported in part by the U.S. Department of Energy,
under grant DE-FG02-91ER40685 and by the U.S. National Science Foundation, 
under grant PHY-9600155.

\vspace{2.0cm}
\noindent {\large \bf Appendix:  Gauge Invariance} \\

Our results include only diagrams with two intermeediate top quarks.
Because these top quarks are not assumed to be on shell, 
our result is not strictly gauge invariant.  We can 
obtain a gauge invariant answer by subtracting some non-doubly
resonant terms as follows.  
Consider the diagram where the gluon is radiated by the top quark
(the first diagram in Fig.~\ref{diagrams}). 
The contribution of this diagram to the production amplitude is
(with $\hat{a}\equiv a^\mu\gamma_\mu$):
\beq
 {\cal{M}}_{prod}^{(t)} \sim \frac{1}{2k p_t}\ \bar{u}(b)\ \hat{\epsilon}_{W}\
 \frac{\hat{p}_{Wb} + m_t}{p_{Wb}^2 - \bar{m}_t^2}\
\hat{\epsilon}_{g} \ (\hat{p}_t+\hat{k} + m_t) \ldots v(\bar{b})\; .
\eeq
By commuting $\hat{\epsilon}_{g}$ to the right, this can be written 
\beq
 {\cal{M}}_{prod}^{(t)} \sim \frac{1}{2k p_t}\ \bar{u}(b)\ \hat{\epsilon}_{W}\
 \frac{\hat{p}_{Wb} + m_t}{p_{Wb}^2 - \bar{m}_t^2}\
( 2 \epsilon_{g} \cdot p_{Wb} + \hat{\epsilon}_{g} \hat{k} +
(p_{Wb}^2 - m_t^2) \hat{\epsilon}_{g} ) \ldots v(\bar{b})\; .
\eeq                     
The term which breaks gauge invariance here is 
the one proportional to $( p_{Wb}^2-m_t^2 )$. 
This is a non-resonant term, regardless of the gluon being
radiated in production or decay stage (in other words, regardless of
$p_{Wb}^2 \approx m_t^2$ or $p_{Wbg}^2 \approx m_t^2 $); therefore, in 
keeping
with the approximation used, we can neglect it. 

A similar analysis works for the contribution of this diagram to the top decay 
 amplitude, with $p_{Wb}$ replaced by  $p_{Wbg}$
in this case, and we drop a term proportional to $( p_{Wbg}^2-m_t^2 )$.
Finally, the amplitudes corresponding  to the diagram in which the gluon
originates from the $\bar{t}$ can be computed in the same manner. The 
final result is gauge invariant, and differs from the exact
result by non-doubly-resonant terms, as we have shown.

We have implemented the above computation in the Monte Carlo program, and 
have checked numerically that the difference between the gauge invariant result
and the exact result is very small (of order 0.01\% of the total cross 
section,  and 
order 1\% with respect to the interference terms). This indicates that the 
other non-doubly resonant contributions (coming from diagrams with a 
single top or none) are also small; a more detailed study is in progress.

Finally, we note  that this method for restoring gauge invariance
is not unique. We could, for example, have instead  replaced the
top mass in the top propagator numerator with the invariant masses:
$\sqrt{p_{Wb}^2}$ in the production amplitude, and  $\sqrt{p_{Wbg}^2}$ in 
the decay amplitudes. The result is also gauge invariant, and also 
differs
from the exact result by non-resonant terms.


\vskip 1truecm
\begin{thebibliography}{99}

\bibitem{lifetime} I.I.~Bigi, {\it et al.}, 
Y.L.~Dokshitzer, V.~Khoze, J.~Kuhn and P.~Zerwas,
Phys.\ Lett.\  {\bf B181}, 157 (1986); L.H.~Orr and J.L.~Rosner,
Phys.\ Lett.\  {\bf B246}, 221 (1990), {\bf 248} (1990)
474(E); L.H.~Orr, Phys.\ Rev.\  {\bf D44}, 88 (1991).

\bibitem{prod}
J.~Jersak, E.~Laerman, and P.~Zerwas, Phys.\ Rev.\
 {\bf D25} 1218 (1982); 
Yu.L.~Dokshitzer, V.A.~Khoze, and W.J.~Stirling, 
Nucl.\ Phys.\  {\bf B428} 3 (1994). 

\bibitem{decay}
M.~Jezabek and J.~H.~Kuhn, Nucl.\ Phys.\  {\bf B314}, 1 (1989);
A.~Czarnecki, Phys.\ Lett.\ {\bf B252}, 467 (1990);
C.~S.~Li, R.~J.~Oakes and T.~C.~Yuan, Phys.\ Rev.\  {\bf D43}, 3759 (1991).


\bibitem{schmidt}C. R. Schmidt, \textit{Phys. Rev.}
{\bf D54} 3250 (1996) 

\bibitem{jikia}
G.~Jikia, Phys.\ Lett.\ {\bf 257B} (1991) 196. 

\bibitem{kos} V.A.~Khoze, L.H.~Orr, and W.J.~Stirling, 
 Nucl.\ Phys.\ {\bf B378} (1992) 413.

\bibitem{mcos} C.~Macesanu, Linear Collider Workshop 2000,
Fermilab, Batavia, IL, Oct.~24--28, 2000, to be published in the 
proceedings, and in preparation.

\bibitem{ks}
R.~Kleiss and W.J.~Stirling, Nucl.\ Phys.\ {\bf B262} (1985) 235.

\bibitem{WW}
A.~Denner, S.~Dittmaier, M.~Roth and D.~Wackeroth,
Nucl.\ Phys.\  {\bf B587}, 67 (2000)

\bibitem{siopsis}
G.~Siopsis,
Phys.\ Rev.\  {\bf D58}, 014009 (1998).


\end{thebibliography}
\end{document}